\documentclass{ws-procs975x65}
\newcommand{\half}{\frac{1}{2}}

\newcommand{\TP}{\stackrel{\leftrightarrow}{\partial}}
\newcommand{\BV}{\left(\begin{array}{c}}
\newcommand{\EV}{\end{array}\right)}
\newcommand{\BM}{\left(\begin{array}{rrrr}}
\newcommand{\BcM}{\left(\begin{array}{cccc}}

\newcommand{\RA}{\rightarrow}

\begin{document}

\title{STERILE NEUTRINOS IN A 6X6 MATRIX APPROACH}

\author{T. GOLDMAN$^*$}

\address{Theoretical Division, Los Alamos National Laboratory,\\
Los Alamos, New Mexico 87545, USA\\
$^*$E-mail: tgoldman@lanl.gov\\
www.lanl.gov}

\begin{abstract}
Quark-lepton symmetry invites consideration of the existence of sterile 
neutrinos.  Long ago, we showed that this approach predicts large 
neutrino mixing amplitudes. Using a Weyl spinor approach, we show, 
in an analytic example, how this, and pseudo-Dirac pairing, can develop 
within a reduced rank version of the conventional see-saw mechanism, 
from small intrinsic mixing strengths. We show by numerical examples 
that mixing of active and sterile neutrinos can affect the structure of 
oscillations relevant to extraction of neutrino mixing parameters from 
neutrino oscillation data.
\end{abstract}

\vspace*{-3.2in}
\begin{flushright} LA-UR-07-4635 \end{flushright}
\vspace*{3.1in}

\keywords{McKellar; Neutrinos; Sterile; Mass; Festschrift.}

\bodymatter

\section{Quark-Lepton Symmetry Is Our Basis}\label{qlsym:sec1}

In the Standard Model (SM) as first formulated, there were no neutrino 
mass terms as no right-chiral projections of Dirac neutrino fields were 
known to exist. Excluding them left no means to produce Dirac neutrino 
mass terms and Majorana mass terms required introduction of either 
non-renormalizable terms in the Lagrangian or new scalar fields with 
unit weak isospin. 

However, the formulation of Grand Unified Theories (GUTs) in the mid-70's,  
$SU(5)$ in particular\cite{GG}, made clear that the fundamental degrees of 
freedom were not Dirac bispinors but two-component Weyl spinors. Later 
developments in supersymmetry and supergravity amplified this contention. 
In the Weyl spinor basis, all known fermions, except the neutrinos, appeared 
in left- and right-chiral pairs ($(\half,0)$ and $(0,\half)$ irreps under the Lorentz 
group). The pairing, along with equal mass terms, were necessary to allow 
construction of Dirac bispinors which could satisfy the known (to high accuracy) 
conservation of electric and color charges. Thus, right-chiral partners for the 
known neutrinos were not required, but, to some of us, at least, seemed strongly 
invited, especially as the successes of quark-lepton symmetry grew over the 
following decade: charm\cite{RT}, and then after the discovery\cite{Perl} of the 
$\tau$-lepton, the $b$-quark\cite{Leon} and eventually the $t$-quark\cite{fnal}. 

\section{See-Saw Mechanism}\label{seesaw:sec2}
At Los Alamos, a number of researchers and visitors, including Stephenson, 
Slansky, Ramond and Gell-Mann,\cite{GMRS} recognized that the right-chiral fields 
were unconstrained, even in $SU(5)$, in the possible Majorana (or as we 
prefer to say here, Weyl) mass term possible -- the mass could be as large as 
the GUT scale ($M~\sim~10^{16}$~GeV). Furthermore, this, combined with now 
normal (order quark or charged lepton) Dirac mass terms ($m~\sim~10^{\pm 3}$~GeV) 
that should appear, would produce eigenstates with very small Majorana masses 
($\sim m^{2}/M$) and that were almost purely left-chiral neutrinos, that is, those 
that participate in the weak interactions. As a bonus, the known 
Cabibbo-Kobayashi-Maskawa (CKM) mixing\cite{ckm} between quark mass and 
weak interaction eigenstates strongly suggested that similar physics should develop 
in the lepton sector, producing the long-conjectured oscillation of neutrinos\cite{ponte} 
(although between different flavors rather than between particle and antiparticle as 
originally suggested). 

\begin{figure}[h]
\psfig{file=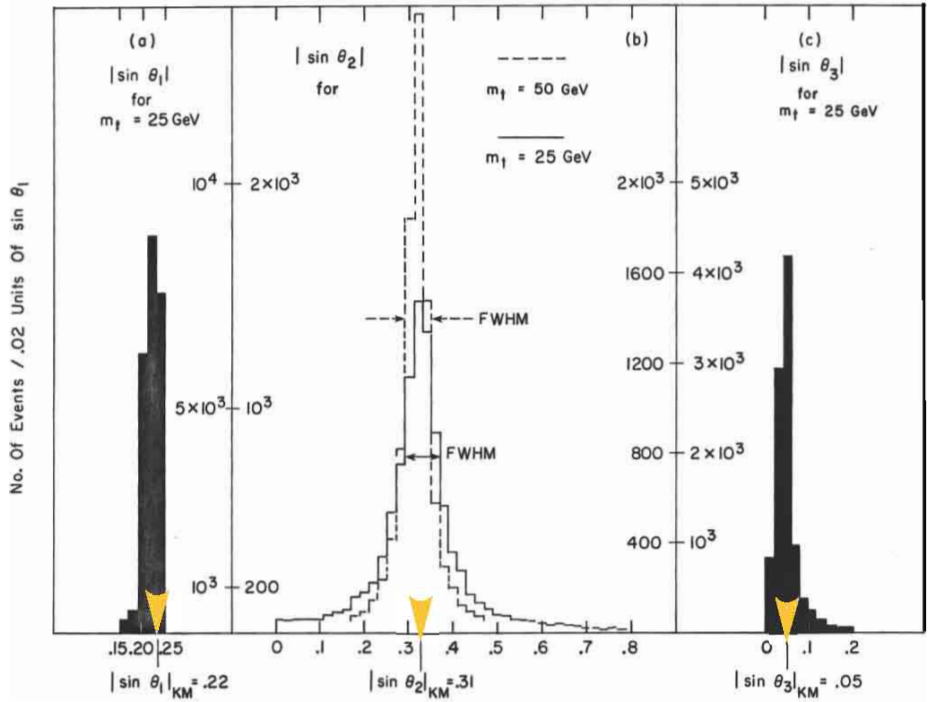,width=5in}
\caption{Mixing angle distributions for random mass matrix entries from Ref.~\refcite{GS}.}
\label{early:fig1}
\end{figure}

\subsection{An Early Effort}\label{early:sec2a}

In the absence of any credible detailed conjectures as to the structure of the 
mass matrices in the lepton sector (although there were a plethora of papers 
about what would now be called "textures"), we\cite{GS} carried out a Monte 
Carlo study (popular more recently in considering the possible values of the 
multiple parameters in supersymmetric theories) using random choices for 
mass matrix entries, allowing for CKM-like mixing of the quarks in the Dirac 
mass sector of the leptons. 

The results were rather astonishing. As Fig.~\ref{early:fig1}, taken from that work shows, 
Cabibbo mixing is favored between the first two "generations" and even larger mixing 
is highly probable between the second two, depending upon how extreme the third 
generation differs in mass. (Note that we did not have the temerity to consider an 
extreme as radical as actually occurs in the quark sector.) The mixing between the 
first and third "generations" is smaller, but non-vanishing. At the time, the only 
potential evidence for neutrino mixing was the intermediate result of the Davis 
experiment\cite{RayD}, which was considered highly suspect, although it was later 
confirmed quite precisely.\cite{sno} We used our result to support mounting of neutrino 
oscillation experiments, saying that the mixing might well be large, although we could 
not predict the scale of the oscillation length. 

\section{Weyl spinors}\label{weyl:sec3}

Since active neutrinos have only two basic states (as opposed to the four of a Dirac 
bispinor), they can be efficiently described in terms of Weyl spinors. We present the 
Lagrangian, equations of motion, and solutions for massive Weyl spinors, then show 
the relation to Majorana and Dirac constructs.

\subsection{Lagrangian Density for Massive Weyl Spinors}\label{lgn:sec3a}

Let the Grassman-valued field variable, $\phi$, represent a left-chiral $(\half,0)$
irrep of the Lorentz Group. Then 
\begin{equation}
{\cal L}_L  =   \half \phi^{\dag} \sigma^\mu \TP_\mu \phi 
  +  \half im \left(\phi^T \sigma^2 \phi + \phi^{\dag} \sigma^2 \phi^* \right) \label{lagl:eq1}
\end{equation}
where
$
\TP \equiv  {\stackrel{\rightarrow}{\partial}} - {\stackrel{\leftarrow}{\partial}}
 $
 and  
$
\sigma^{\mu} = (1, \sigma^i)
$
with $\sigma^i$ the Pauli matrices. 
Under a Lorentz transformation with parameters, $\omega_{\mu}$, 
\begin{equation}
\phi \RA e^{-\frac{\imath}{2} (\sigma^{\mu} \omega_{\mu})} \phi \label{lftbst:eq2}
\end{equation}

For a right-chiral $(0, \half)$ irrep of the Lorentz Group, we need only 
make the substitution:
\begin{equation}
{\cal L}_R :  \sigma^\mu  \RA  \bar{\sigma}^\mu = (1, -\sigma^i) 
\end{equation}
in Eq.~\ref{lagl:eq1} to acquire the relevant Lagrangian.

Because mass terms must couple left-chiral $(\half,0)$ and right-chiral 
$(0, \half)$ irreps of the Lorentz Group, it is apparent that  
\begin{equation}
\chi = \sigma^2 \phi^* \label{rhsp:eq4}
\end{equation}
must be in a right-chiral $(0, \half)$ irrep, as can be seen by applying the
boost in Eq.~\ref{lftbst:eq2} to $\phi$ irrep and then applying the commutation 
rules for the Pauli matrices to the construction in Eq.~\ref{rhsp:eq4}. This will 
be relevant shortly. 

\subsection{Equations of Motion and Form of Solutions}\label{eom:sec3b}

Writing $\phi$ out explicitly as a two component column spinor, 
\begin{equation}
\phi = 
\BV
\phi_{1} \\ \phi_{2} 
\EV
\end{equation}
the equations of motion for the components become
\begin{eqnarray}
\partial_t \phi_{1} - \partial_z \phi_{1} -(\partial_x -\imath  \partial_y ) \phi_{2} & = & -m \phi_{2}^{*}  \nonumber \\
\partial_t \phi_{2} +\partial_z \phi_{2} -(\partial_x +\imath  \partial_y ) \phi_{1} & = & + m \phi_{1}^{*} 
\end{eqnarray}

Defining $\theta = Et - \vec{p}\cdot\vec{x}$ and $p_{\pm} = p_{x} \pm \imath p_{y}$,
we find the complex conjugate pair of solutions, $\phi_{-}$ and $\phi_{+} = \phi_{-}^{*}$  
to have the form
\begin{equation}
\phi_{-} =  \BV
F e^{-\imath \theta} \\ 
-\frac{p_{+}}{E-p_{z}}F e^{-\imath \theta} -\imath \frac{m}{E-p_{z}}F^{*} e^{+\imath \theta} \label{solnl:eq7}
\EV
\end{equation}
where $F$ is a Grassman-valued constant. 

\subsection{Majorana and Dirac Bispinors}\label{mdbsp:sec3c}

A Majorana bispinor is simply a redundant representation of the Weyl spinor above. 
In the Wigner-Weyl representation for the bispinor, we make use of the transformation 
in Eq.~\ref{rhsp:eq4}, to construct
\begin{equation}
\Psi_{M} =  
\BV
\phi \\ 
e^{\imath \eta} \sigma^2 \phi^* \label{ww} 
\EV
\end{equation}
where the phase $\eta$ can be chosen as $0$, $\pm \pi/2$ or $\pi$ for later convenience. 
The field $\Psi_{M}$ has a Lagrangian that can be put into Dirac form with mass $m$. 

To construct a Dirac bispinor, two independent $(\half,0)$ irreps must be invoked, 
which we labe suggestively as $a$, or active neutrino in the SM and $s$, for 
sterile neutrino in the SM. (Except for the $U(1)$ factor, these terms apply to the 
left- and right-chiral parts of the charged fermions as well.) Thus, 
\begin{equation}
\Psi_{D} =  
\BV
a \\ 
- \sigma^2 a^*
\EV 
+ \imath
\BV
s \\ 
- \sigma^2s^*
\EV
= \Psi_{a} + \imath \Psi_{s}
\end{equation}
where the phase choices have been made so that if $\Psi_{a}$ and $\Psi_{s}$ have 
the same mass value $m$, then (as can be seen from Eq.~\ref{lagl:eq1}) $\imath\Psi_{s}$ 
has mass value $-m$ and a $45^{\circ}$ rotation in the basis space will explicitly display 
$m$ as a Dirac mass. (See Eq.~\ref{ww2pd:eq10} below.)

The rest states of two such independent spinors (see Eq.~\ref{solnl:eq7}), with independent 
Grassman constants $F$ and $G$, can be combined (with $F = -G$ and $H$ the sum) to 
produce the familiar form of a spin-up Dirac particle in the Pauli-Dirac representation, {\it viz.} 
\begin{equation}
\frac{1}{\sqrt{2}} \BM
\;1 & \;0 & \; 1 & \;0 \\
\;0 & \;1 & \;0 & \;1 \\
-1 & \;0 & \;1 & \;0 \\
\;0 & -1 & \;0 & \;1 
\EV
\BV
H e^{-\imath m t}  \\ 
0 \\
H e^{-\imath m t}  \\ 
0
\EV
= 
\BV
A e^{-\imath m t} \\
0 \\ 0 \\ 0 \EV \label{ww2pd:eq10}
\end{equation}
where again, $F, G, H$ and $A$ are all Grassman-valued constants. 

\section{Beyond the Simple See-Saw: Reduced Rank}\label{redrnk:sec4}

The see-saw mechanism, as originally invoked, assumed, for simplicity, 
that the Majorana mass matrix of the right-chiral fields, represented as 
left-chiral but sterile neutrinos, is proportional to the unit matrix. Since 
then, many different "textures" for mass matrices of the fundamental fermions 
have been conjectured. We\cite{SMG} (and others\cite{Mohap}) have studied 
the possibility that the rank of this so-called right-handed mass matrix is less 
than three. 

With quark-lepton symmetry, the mass matrix structure consists of four 
three-by-three blocks: The $(1,1)$ block describes the Majorana masses 
of the active neutrinos and must vanish in the absence of a triplet Higgs 
field. The $(1,2)$ and $(2,1)$ blocks describe Dirac mass terms ($m$) that 
couple the active and sterile Weyl spinor neutrino fields. If we take them to 
be diagonal for the moment, this defines the flavor of each sterile neutrino 
field as a partners of a particular active neutrino. Finally, the $(2,2)$ block 
describes the Majorana masses of the sterile neutrinos. In this basis, we can 
initially, for a rank one sterile mass matrix, set all of the entries to zero except 
for the $(3,3)$ element of the$(2,2)$ block, which we label $M$. 

These alignments are unrealistic, of course, so we carry out two sets of rotations: 
One corresponds to moving the vector $(0, 0, M)$ in the sterile "flavor" space 
away from the "3" axis with the standard angles, $\theta$ (from the 3-axis) and 
$\phi$ (in the 1-2 plane). In addition, we allow for CKM-like mixing in the Dirac 
mass matrix sector. Absent CP-violation, this accounts for all of the possible 
mixings among the six fields and their corresponding particle states. 

We found in this system, that there is a wide range of parameters over which 
the mass eigenstates form into a very light, mostly active neutrino as in the 
conventional see-saw, a very heavy (under the assumption that $M >> m$) 
mostly sterile neutrino, again conventionally, and two pairs of "pseudo-Dirac" 
neutrinos, the the sense of Wolfenstein\cite{wolf}. The resulting mixing 
amplitudes among the active neutrinos are very large to maximal. 

\subsection{Analytic Analysis of Two Flavor Case}\label{2flvr:4a}

In order to understand this better, we solve analytically the simpler two 
flavor case. The eigen equation for this system is, after rotation from 
exact alignment (and ignoring the analog of CKM mixing), 
\begin{equation}
\mu_{i} \Phi_{i} = 
\BcM
\;0 & \;0 & m_{1} & \;0 \\
\;0 & \;0 & \;0 & m_{3} \\
m_{1} & \;0 & Ms^{2} & Mcs \\
\;0 & m_{3} & Mcs & Mc^{2} 
\EV
\BV
\alpha_{i}  \\ 
\beta_{i}  \\
\gamma_{i}   \\ 
\delta_{i} 
\EV
\end{equation}
where the $\Phi_{i}$ are four component column vectors with entries as indicated 
on the RHS of the equation, and $s = sin\theta$ and $c = cos\theta$ where 
$\theta$ is the misalignment angle between active and sterile flavor spaces. 

\begin{figure}[b]
\psfig{file=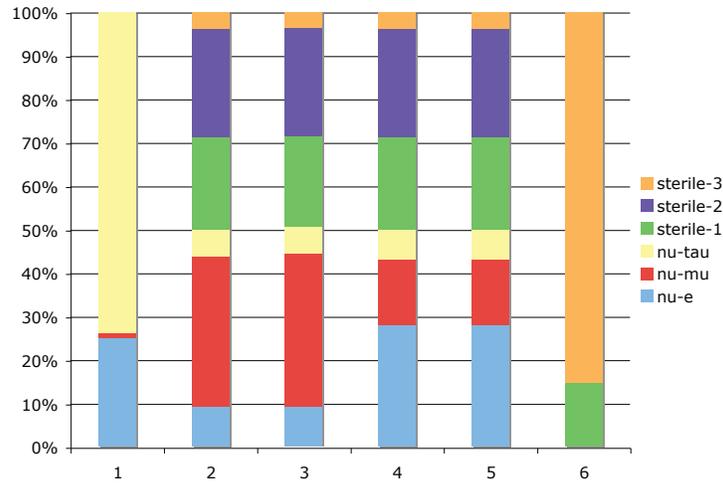,width=5in}
\caption{Three flavor example with parameters chosen to demonstrate large flavor 
mixing and two pseudo-Dirac pairs ([2,3] and [4,5]). The bands indicate the relative 
amplitudes of each active and sterile flavor in each eigenmass state enumerated 
along the baseline.}
\label{2pds:fig2}
\end{figure}

McKellar showed that the eigenvalues of this system are
\begin{eqnarray}
\mu_{1} & = & +m_{0} - \frac{a^{2}}{M}  - \frac{a^{2}}{m_{0}M^{2}}( m_{0}^{2} - \frac{a^{2}}{2} -b^{2})  \nonumber \\
\mu_{2} & = &  -m_{0} - \frac{a^{2}}{M}  + \frac{a^{2}}{m_{0}M^{2}}( m_{0}^{2} - \frac{a^{2}}{2} -b^{2}) \nonumber \\
\mu_{3} & = &  - \frac{b^{2}}{M} + \mathcal{O}(M^{-3})  \nonumber \\
\mu_{4} & = & M + \frac{2a^{2}+b^{2}}{M} + \mathcal{O}(M^{-3})
\end{eqnarray}
where 
\begin{eqnarray}
m_{0}^{2} &  = & m_{1}^{2} \;{\rm cos}^{2}\theta  + m_{3}^{2}\; {\rm sin}^{2}\theta \nonumber \\
a &  = & \frac{(m_{1}^{2}-m_{3}^{2})\;{\rm cos}\theta\; {\rm sin}\theta}{m_{0}\sqrt{2}} \nonumber \\
b &  = & \frac{m_{1}m_{3}}{m_{0}}
\end{eqnarray}
Note the low and high mass see-saw pair, $\mu_{3}$ and $\mu_{4}$, and the psuedo-Dirac 
pair, $\mu_{1}$ and $\mu_{2}$, which would form a single Dirac neutrino to this order if it 
happened that $a = 0$. 
 
\def\figsubcap#1{\par\noindent\centering\footnotesize(#1)}
\begin{figure}[b]
\begin{center}
 \parbox{2.1in}{\epsfig{figure=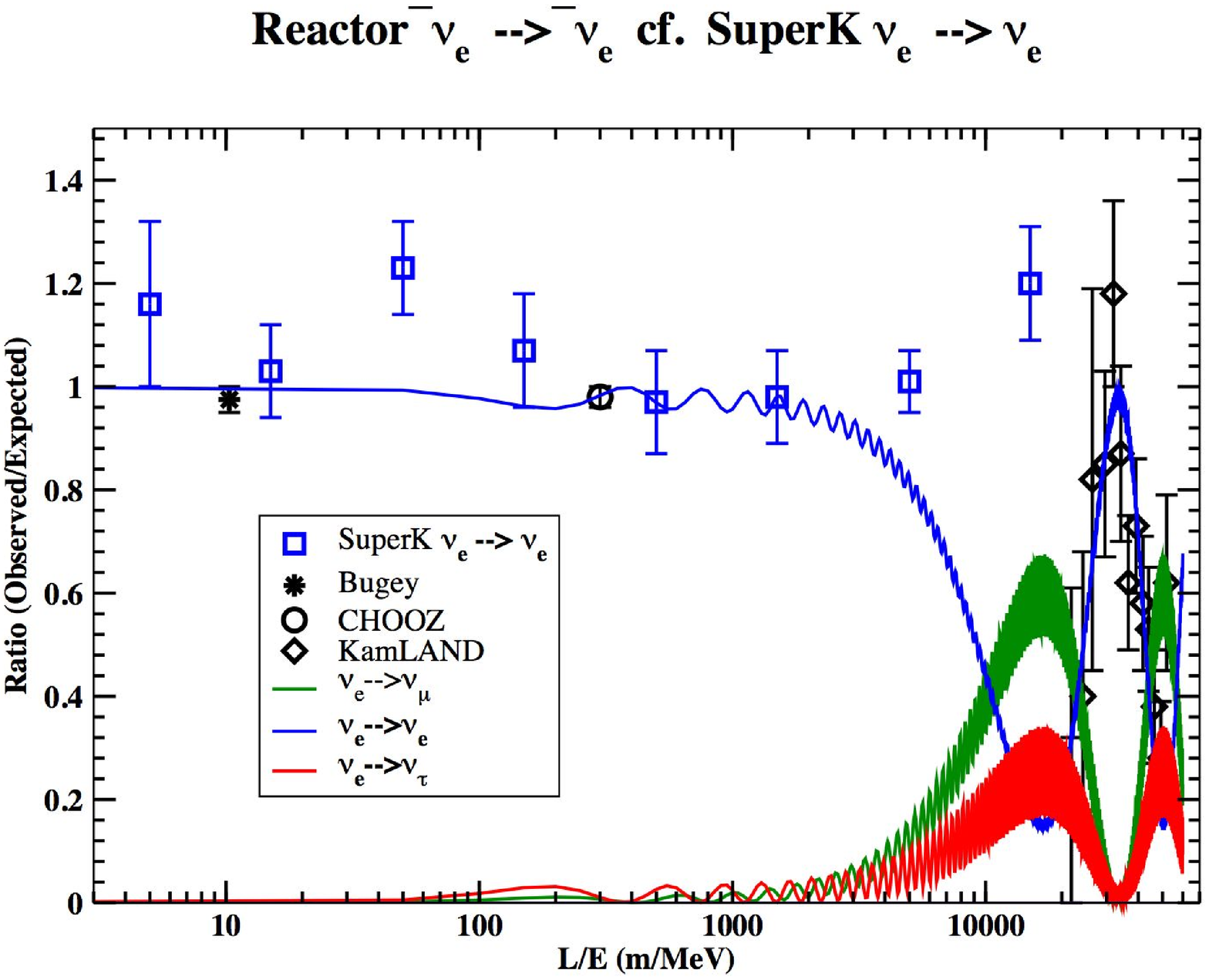,width=2.5in}
 \figsubcap{a}}
 \hspace*{4pt}
 \parbox{2.1in}{\epsfig{figure=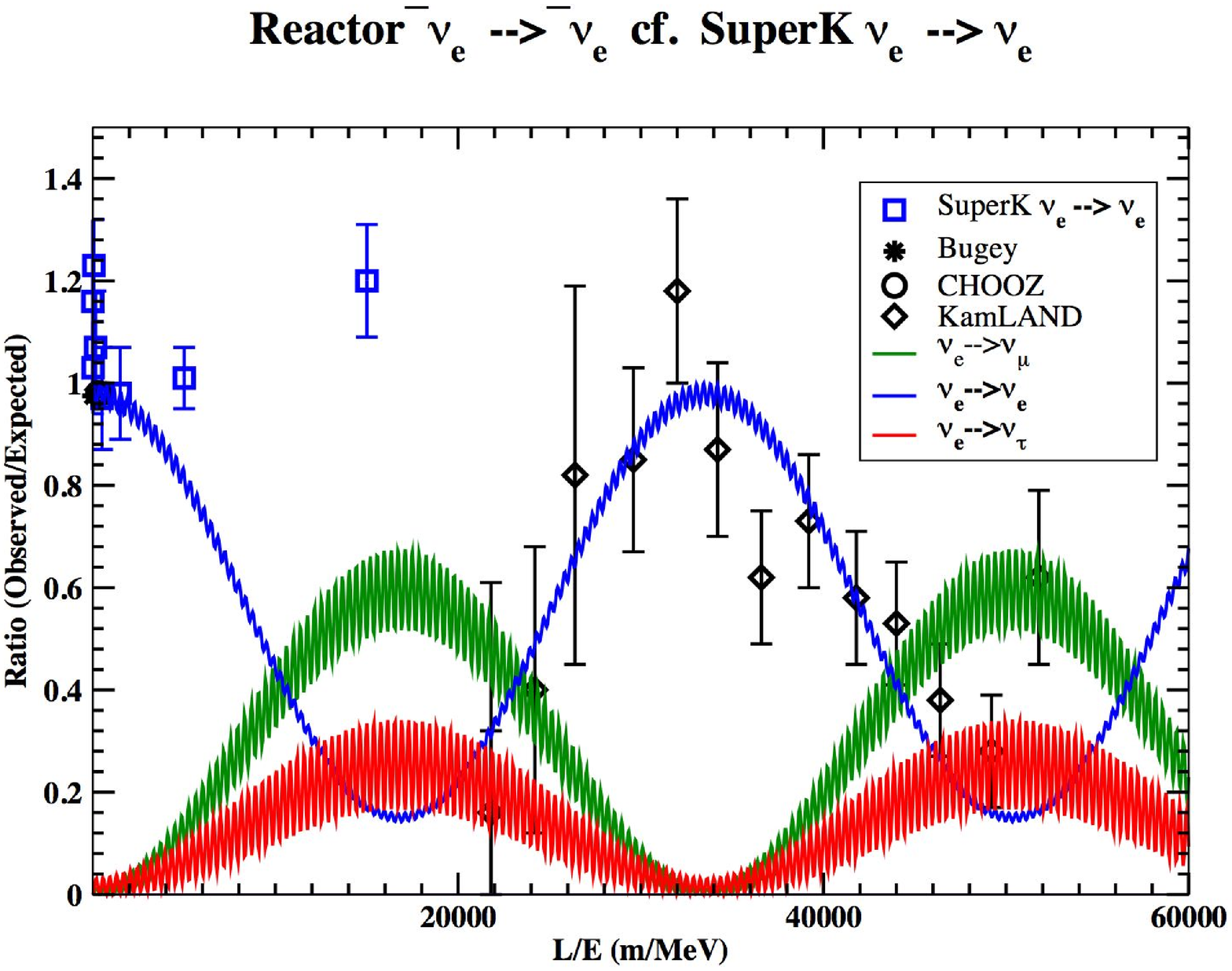,width=2.5in}
 \figsubcap{b}}
 \caption{Electron antineutrino disappearance vs. ratio of distance from 
source divided by neutrino energy compared with experimental data.
 (a) Logarithmic plot. (b) Linear plot.}
\label{kam:fig3}
\end{center}
\end{figure}
 
The eigenvector components for these solutions satisfy
\begin{equation}
\frac{\alpha_{1}}{\beta_{1}} =  \frac{\alpha_{2}}{\beta_{2}}  = 
\frac{\beta_{3}}{\alpha_{3}}  = \frac{m_{1}}{m_{3}} {\rm cot}\theta \label{lgmx:eq14}
\end{equation}
and 
\begin{equation}
\frac{\gamma_{i}}{\alpha_{i}} =  \frac{\mu_{i}}{m_{1}}  \;\;\; ; \;\;\; \frac{\delta_{i}}{\beta_{i}} =  \frac{\mu_{i}}{m_{3}}
\end{equation}
Hence for relatively large $M$ and small $\theta$, eigenstates $3$ and $4$ are almost purely 
active and sterile respectively, while $\mu_{1} \sim \mu_{2} \sim m_{1}$ and the states with 
these two eigenvalues will have large components of both active flavor states when 
\[
\frac{m_{3}}{m_{1}} \sim {\rm cot}\theta
\]
that is, the terms in Eq.~\ref{lgmx:eq14} are of order one. 

\subsection{Results for Three Flavor Case} \label{sec4b}

An analytic demonstration is not so easy to provide in the three flavor case, 
but a similar result, with two pseudo-Dirac pairs and large flavor mixing, is 
shown in Fig.~\ref{2pds:fig2} as an example case. 

\begin{figure}[b]
\psfig{file=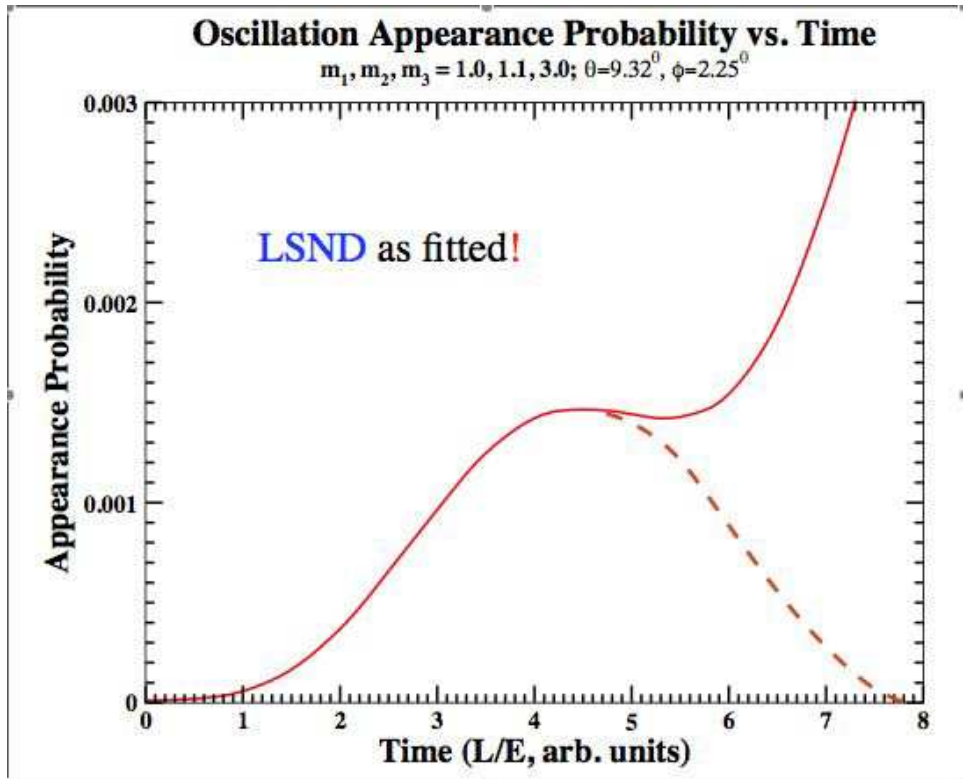,width=5in}
\caption{Comparison of electron neutrino appearance vs. ratio of distance from source 
over neutrino energy as predicted by our lower rank, six-channel mixing model and the 
functional form (dashed line) assumed in the fit made by the experimental group.}
\label{lsnd:fig4}
\end{figure}

In Fig.~\ref{kam:fig3}(a), we show the result of a specific choice of parameters for electron 
antineutrino disappearance, which is consistent with the results of atmospheric and 
reactor experiments. Within uncertainties, the last two high points of the atmospheric 
(SuperKamiokande\cite{sk}) experiment are consistent in our parametrization with 
feed-through into electron neutrinos from the disappearance of muon neutrinos, also 
seen in that experiment. Fig.~\ref{kam:fig3}(b) focuses in detail on the region studied by 
the KamLAND experiment.\cite{kam}

In Fig.~\ref{lsnd:fig4}, we show the shape of the electron neutrino appearance function 
appropriate to the LSND experiment\cite{lsnd} and contrast this with the shape of 
the fitting function actually used (with the dashed extension corresponding to the simple 
sinusoidal function of two-channel mixing). This demonstrates that incorrect parameters 
may be extracted from experiments by not fitting directly to the full panoply of possibilities 
allowed by the three known flavors of neutrinos. 

\begin{figure}[b]
\psfig{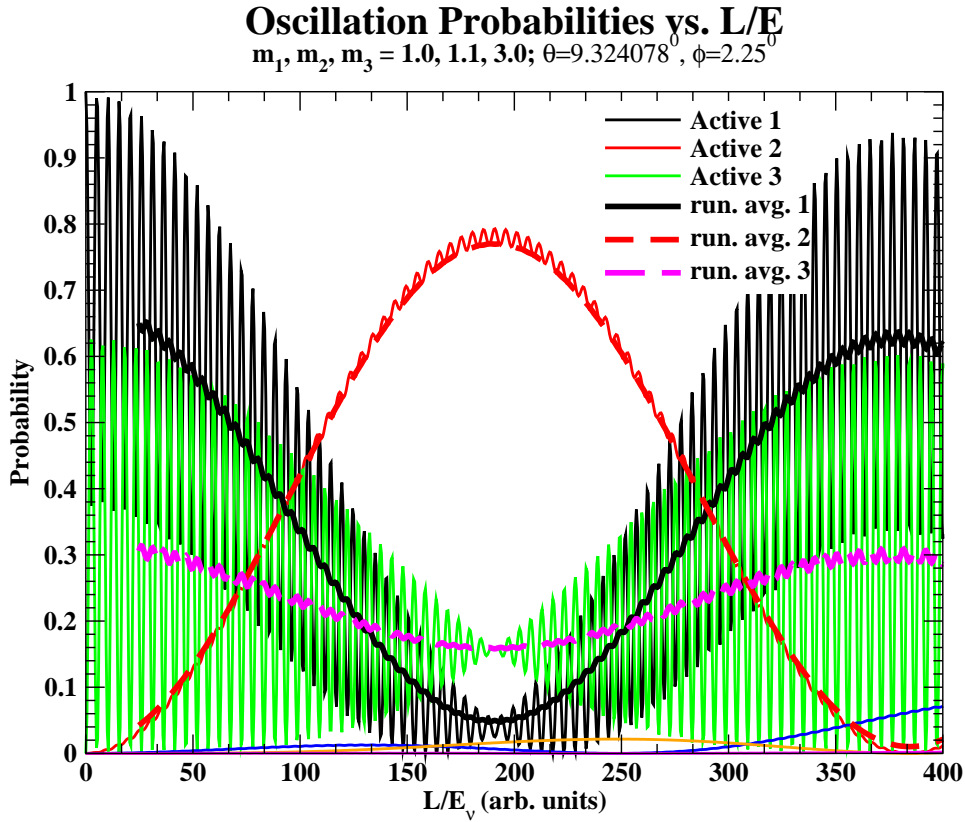}
\caption{Effect of finite resolution on disappearance and appearance as a function of time 
in the neutrino rest frame (or equivalently, ratio of distance from source to neutrino energy). 
The average of the disappearance of the initial neutrino flavor has approximately the expected 
shape for a much longer wavelength mixing than is seen at high resolution. See text for more 
discussion.}
\label{runavs:fig5}
\end{figure}

Finally, in Fig.~\ref{runavs:fig5}, we show an example of how finite resolution, particularly 
in the neutrino energy, affects the oscillation pattern that is observed. The rapid oscillation 
between the initial muon neutrino and strongly mixed tau neutrino is smeared into a much 
longer wavelength average muon neutrino disappearance. Without explicit detection of 
tau neutrinos, it is difficult to discern that their appearance is not following the expected, 
simple sinusoidal  two channel appearance structure, but rather is almost constant, falling 
slightly while electron neutrinos actually appear in a growing fashion. Fig.~\ref{lsnd:fig4} 
is a magnification of a very small region near the origin of this plot , and it is with the 
parameters noted here that the exceptional points referred to in Fig.~\ref{kam:fig3} are 
explained. 

\begin{figure}[t]
\psfig{file=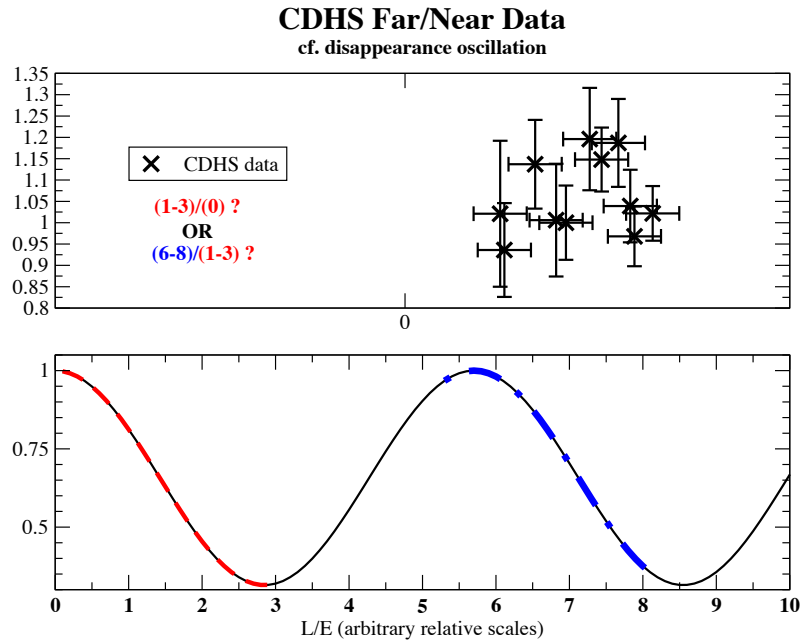,width=5in}
\caption{Illustration of how the distance-corrected ratio of events in a far detector to those in 
a near detector can exceed unity if a very short wavelength oscillation affects he rate in the 
near detector and comparison with CDHS results.\cite{cdhs}. See text for more 
discussion.}
\label{cdhs:fig6}
\end{figure}

\subsection{One Detector or Two?}\label{numdets:sec4c}

A final note in passing: A vigorous debate that continues in the experimental community 
concerns the question of whether spectral distortion or two detectors at different distances 
from the same neutrino source affords the better means to observe and study neutrino 
oscillation phenomena. The latter is, of course, generally more expensive, so one might 
be inclined to think it is also more valuable. A curious result from the CDHS experiment, 
however, demonstrates that one must first be certain that the near detector is so close that 
no oscillations at all have taken place by the time the beam arrives at that detector. In the 
CDHS results reported\cite{cdhs}, the flux in the farther detector is generally larger than in 
the near detector. Since this violates unitarity, it allows for a very stringent limit on neutrino 
disappearance. However, as we illustrate in Fig.~\ref{cdhs:fig6}, the excess can also be due to 
the near detector reacting to the first wave of oscillation disappearance, with the far 
detector appearing to have a larger ($L^{2}$ corrected) flux by being at a slightly different 
phase in the oscillation wave. 

\section{Conclusion} \label{concl:sec5}

We draw several conclusions from the above, not all of them warranted. 
\begin{itemize}
\item There may well be 5 independent neutrino mass differences that 
must be fit to neutrino oscillation experimental data. 
\item Analyses of oscillation data in terms of $2 \times 2$ mixing can miss 
significant physics and even lead to extraction of invalid parameter values. 
\item A global, multichannel analysis is essential before firm conclusions 
can be reliably drawn regarding neutrino masses and mixing parameters. 
\item With the $\sim$eV mass scales we have examined, neutrinos can 
contribute significantly to the Dark Matter in the Universe. 
\end{itemize}

This paper includes work done with Jerry Stephenson and Bruce McKellar 
over many years. It has been our great pleasure to work with Bruce and we 
hope that, as soon as he retires, he will have a lot more time to work with us! 

This work was carried out under the auspices of the National Nuclear Security 
Administration of the U.S. Department of Energy at Los Alamos National Laboratory 
under Contract No. DE-AC52-06NA25396.

\bibliographystyle{ws-procs975x65}
\bibliography{ws-pro-sample}

\end{document}